\documentclass[epj]{svjour}
\usepackage{graphicx}

\usepackage{amssymb}
\usepackage{amsmath}

\usepackage[ps2pdf,colorlinks]{hyperref}

\AtBeginDocument{%
    \hypersetup{
      urlcolor     = blue, 
      linkcolor    = black, 
      citecolor   = blue, 
      pdfborder={0 0 0},
      urlbordercolor={1 1 1},
      citebordercolor={1 1 1}
    }
    }

\begin{document}
\title{Field operator transformations in Quantum Optics using a novel graphical method with applications to beam splitters and interferometers}
\author{Stefan~Ataman
}                     
%
%
\institute{ECE Paris, \email{ataman@ece.fr}}
\date{Received: date / Revised version: date}
%
\abstract{In this paper we describe a novel, graphical method,
allowing the fast computation of field operator transformations
for linear lossless optical devices in Quantum Optics (QO). The
advantage of this method grows with the complexity of the
considered optical setup. As case studies we examine the field
operator transformations for the beam splitter (BS), the
Mach-Zehnder interferometer (MZI) and the double MZI. We consider
the simple case with monochromatic input light, as well as
extensions to the non-monochromatic case.
%
} 

\titlerunning{Field operator transformations in Quantum Optics using a novel graphical method}
\authorrunning{Stefan Ataman}

\maketitle
%

\section{Introduction}


The quantum optical description of simple optical systems
\cite{Leo03} allows one to compute and predict the behavior of
both classical and non-classical states of light. The beam
splitter (BS) is one of the most widely used devices in optical
experiments. In the classical description of a lossless BS, energy
conservation imposes the relation between input and output
electric fields \cite{Lou03,Gry10} (also known as the Stokes
relations). In the quantum optical description
\cite{Yur86_Cam89,Fea87,Pra87}, fields are replaced by operators. 
Starting from Schwinger's work on the angular momentum operators,
general frameworks have been developed, where beam-splitters are
described in SU(2) symmetries \cite{Yur86_Cam89}. However, authors
typically focus on a subset of this general model, having
symmetrical \cite{Lou03,Ger04,Fea87} or non-symmetrical operator
input-output operator relations \cite{Gry10,Man95}.

Beam splitters proved to be pivotal in experiments performed in
QO, for example by helping differentiate between a coherent and a
Fock state \cite{Gra86,Asp91} and reveal non-classical features of
light \cite{Kim77_Gho87,HOM87_Kim00_Pit96_Kim03}. Applying a
coherent state \cite{Gla63,Sud63,Gla63b} at one input and a
single-quantum Fock state at the other one of a BS allows
measuring quantum states of light using the homodyne detector
\cite{Yue78_Yue83,Leo95}.


A Mach-Zehnder interferometer (MZI) is a device composed of two
beam splitters and two mirrors \cite{Lou03,Man95,Ger04}. Its
versatility has led to its use in countless experiments
\cite{Gra86,Rar90,Fra91,Kwi95}. With a single light quantum at one
input, the rate of photo-detection at its outputs oscillates as
the path-length difference of the interferometer is swept
\cite{Gra86}. 
Applying pairs of light quanta at its inputs, specific
non-classical effects show up \cite{Rar90}.


The description of the output states of a BS with given
(especially non-classical) input states stirred a lot of interest.
Kim \emph{et al.} \cite{Kim02} consider a BS with a variety input
states, conjecturing that an entangled output state needs a
non-classical input state. The proof was given through a theorem
in \cite{Xia02}. The transformation relation of field operators
for beam splitters has been considered in \cite{Ou87}, where a
formal solution is given using the P-representation of coherent
states and the optical equivalence theorem. The two-photon
interference at the output of lossless beam splitters has been
thoroughly discussed by Fearn and Loudon \cite{Fea89}. Campos
\emph{et al.} \cite{Cam90} extend this discussion with a special
emphasis on the fourth-order interference applied to lossless
optical systems (BS and MZI). Other authors studied more specific
scenarios, for example displaced Fock states \cite{Win11} or
output photon statistics for input squeezed light \cite{Pla94}.

The computation of input-output operator relations for optical
devices comprising beam splitters and interferometers is
traditionally composed of a cascade of successive operator
transformations. However, the complexity of these operations grows
with the number of cascaded devices in the system. Moreover, due
to this calculatory complexity, a good deal of physical insight is
lost.

The purpose of this paper is to introduce a graphical method
allowing the fast computation of these field operator
transformations. Moreover, contrary to the traditional iterative
method, all computations remain very intuitive because of the
direct physical meaning that can be attached to them. The input
light is considered in both the monochromatic and in the more
complicated, the non-monochromatic case.


This paper is organized as follows. In Section
\ref{sec:field_op_transf_theoretical} we give a theoretical
motivation for the computation of the field operator
transformations. In Section \ref{sec:graph_method_BS_MZI}, using
the newly introduced graph-based method, field operator
transformations are computed for a beam splitter and a
Mach-Zehnder interferometer. In Section
\ref{sec:graph_method_2MZI} the same transformations are computed
for a double Mach-Zehnder interferometer. Finally, conclusions are
drawn in Section \ref{sec:conclusions}.


\section{State and field operator transformations in Quantum Optics}
\label{sec:field_op_transf_theoretical}

Quite often, interesting devices in QO have two input ports
(\emph{e.g.} beam splitter, MZI). We shall label them with the
indexes $0$ and $1$. For simplification, we also assume two output
ports, labelled $N$ and $N+1$. We assume linear and lossless
optical systems.

If the input is in a pure state, and moreover, if we assume
monochromatic light quanta, we can write the input state vector of
our system as\footnote{Strictly speaking, we should have written
${\vert\psi_{in}\rangle=f(\hat{a}_0^\dagger,\hat{a}_0,\hat{a}_1^\dagger,\hat{a}_1)\vert0\rangle}$.
But since any operator function can be normally ordered
\cite{Sud63} and since the annihilation operator at any power
except zero acting on the vacuum state will simply vanish, we end
up with Eq.~\eqref{eq:psi_in_f_vacuum_monochromatic}. An example
with coherent states follows.}
\begin{equation}
\label{eq:psi_in_f_vacuum_monochromatic}
\vert\psi_{in}\rangle=f\left(\hat{a}_0^\dagger,\hat{a}_1^\dagger\right)\vert0\rangle
\end{equation}
where $f$ is an operator function to be determined,
$\hat{a}_k^\dagger$ is the creation operator for the port $k$
(with $k=0,1$) and $\vert0\rangle$ denotes the vacuum state. For
example, if we input the Fock state
${\vert\psi_{in}\rangle=\vert0_01_1\rangle=\hat{a}_1^\dagger\vert0\rangle}$
we find
${f\left(\hat{a}_0^\dagger,\hat{a}_1^\dagger\right)=\hat{a}_1^\dagger}$.
For a coherent state
${\vert\psi_{in}\rangle=\vert0_0\alpha_1\rangle=\hat{D}_1\left(\alpha\right)\vert0\rangle}$
(where
$\hat{D}_1\left(\alpha\right)=\text{e}^{\alpha\hat{a}_1^\dagger-\alpha^*\hat{a}_1}$
is the displacement operator \cite{Gla63} acting on input port
$1$), after normally ordering we have
\begin{equation}
f\left(\hat{a}_0^\dagger,\hat{a}_1^\dagger\right)
=\text{e}^{-\frac{\vert\alpha\vert^2}{2}}\sum_{k=0}^{\infty}{\frac{1}{k!}\left(\alpha\hat{a}_1^\dagger\right)^k}
\end{equation}
If one wishes to find the output state, the fact that an input
vacuum state transforms into an output vacuum state can always be
used, no matter how complicated the system is. Therefore, if we
could find the operator functions $g_0$ and $g_1$ so that
\begin{equation}
\hat{a}_0^\dagger=g_0\left(\hat{a}_N^\dagger,\hat{a}_{N+1}^\dagger\right)
\end{equation}
and
\begin{equation}
\hat{a}_1^\dagger=g_1\left(\hat{a}_N^\dagger,\hat{a}_{N+1}^\dagger\right)
\end{equation}
then, at least formally, the output state can be written as
\begin{equation}
\label{eq:psi_out_f_vacuum_monochromatic}
\vert\psi_{out}\rangle=f\left(g_0\left(\hat{a}_N^\dagger,\hat{a}_{N+1}^\dagger\right),
g_1\left(\hat{a}_N^\dagger,\hat{a}_{N+1}^\dagger\right)\right)\vert0\rangle
\end{equation}
We may call this formalism ``the Schr\"odinger picture'', since
the state vector evolves \emph{i.e.}
$\vert\psi_{in}\rangle\to\vert\psi_{out}\rangle$. Extension to
density matrices instead of state vectors can be readily done.


In most cases, however, the input light is not (or cannot be
approximated to be) monochromatic. Therefore, the extension to
multi-mode fields is needed. We will be using a narrowband
continuous-mode extension, already considered in the literature
\cite{Cam90,Blo90,Leg03}. We assume a Heisenberg picture with
\begin{equation}
\label{eq:EX_fct_zeta_X_FREQ}
\hat{E}_j^{(+)}\left(\omega\right)=\tilde{\zeta}_j\left(\omega\right)\hat{a}_j
\end{equation}
representing the frequency-domain distribution of the output
positive frequency electric field operator and we denoted
${j=N,N+1}$. We also assume that these operator functions can be
Fourier transformed in order to obtain their time-domain
counterparts $\hat{E}_N^{(+)}\left(t\right)$ and
$\hat{E}_{N+1}^{(+)}\left(t\right)$ (via the Fourier transform) as
\begin{equation}
\label{eq:EX_fct_zeta_X_TIME} \hat{E}^{(+)}_j\left(t\right)
=\frac{1}{\sqrt{2\pi}}\int{\tilde{\zeta}_j\left(\omega\right)\hat{a}_j\text{e}^{i\omega{t}}\text{d}\omega}
=\zeta_j\left(t\right)\hat{a}_j
\end{equation}
where $j=N,N+1$. We also state a result from the Fourier theory
\cite{Coh95} that will be used throughout this paper, namely the
delay theorem. If a given function $\xi\left(t\right)$ has the
Fourier transform $\tilde{\xi}\left(\omega\right)$, then
$\tilde{\xi}\left(\omega\right)\text{e}^{-i\omega{\tau}}$
corresponds in the time-domain to a delayed version
$\xi\left(t-\tau\right)$ \emph{i.e.}
\begin{equation}
\label{eq:Fourier_DELAY_theorem}
\frac{1}{\sqrt{2\pi}}\int{\text{e}^{-i\omega{\tau}}\tilde{\xi}\left(\omega\right)\text{e}^{i\omega{t}}\text{d}\omega}
=\xi\left(t-\tau\right)
\end{equation}
Quite often, the coincidence/singles detection probabilities at
the output ports $N$ and/or $N+1$ are needed. For example, if
ideal photo-detectors are assumed, the singles detection rate at
the output port $N$ between the times $t$ and $t+\text{d}t$ is
given by
\begin{eqnarray}
\label{eq:introd_Singles_det_rate_P_N}
P_N\left(t\right)=\langle\psi_{out}\vert\hat{E}_N^{(-)}\left(t\right)
\hat{E}_N^{(+)}\left(t\right)\vert\psi_{out}\rangle
 \qquad
\nonumber\\
\qquad
=\Vert\hat{E}_N^{(+)}\left(t\right)\vert\psi_{out}\rangle\Vert^2
\end{eqnarray}
where
$\hat{E}_N^{(-)}\left(t\right)=\left[\hat{E}_N^{(+)}\left(t\right)\right]^\dagger$.
If we suppose that the field operator
$\hat{E}_N^{(+)}\left(t\right)$ can be written in respect with the
input field operators as
\begin{equation}
\hat{E}_N^{(+)}\left(t\right)=g_N\left(\hat{E}_0^{(+)}\left(t\right),\hat{E}_1^{(+)}\left(t\right)\right)
\end{equation}
and since we chose the Heisenberg picture where state vectors do
not change, the singles detection rate from
Eq.~\eqref{eq:introd_Singles_det_rate_P_N} can be written as
\begin{equation}
P_N\left(t\right)=\Big\Vert
g_N\left(\hat{E}_0^{(+)}\left(t\right),\hat{E}_1^{(+)}\left(t\right)\right)\vert\psi_{in}\rangle\Big\Vert^2
\end{equation}
Similarly, the coincident detection rate at the output ports reads
\begin{eqnarray}
\label{eq:introd_Coincidence_det_rate_P_c}
P_c\left(t,t+\tau_d\right)=\langle\psi_{out}\vert\hat{E}_N^{(-)}\left(t\right)\hat{E}_{N+1}^{(-)}\left(t+\tau_d\right)
\quad
\nonumber\\
\quad\hat{E}_{N+1}^{(+)}\left(t+\tau_d\right)\hat{E}_N^{(+)}\left(t\right)\vert\psi_{out}\rangle
\end{eqnarray}
and one needs the function $g_{N+1}$ so that
\begin{equation}
\hat{E}_{N+1}^{(+)}\left(t\right)=g_{N+1}\left(\hat{E}_0^{(+)}\left(t\right),\hat{E}_1^{(+)}\left(t\right)\right)
\end{equation}
and therefore the coincidence probability from
Eq.~\eqref{eq:introd_Coincidence_det_rate_P_c} can be formally
written as
\begin{eqnarray}
P_c\left(t,t+\tau_d\right)= \Big\Vert
g_{N+1}\left(\hat{E}_0^{(+)}\left(t+\tau_d\right),\hat{E}_1^{(+)}\left(t+\tau_d\right)\right)
\nonumber\\
g_N\left(\hat{E}_0^{(+)}\left(t\right),\hat{E}_1^{(+)}\left(t\right)\right)\vert\psi_{in}\rangle\Big\Vert^2
\end{eqnarray}
The interest in finding the operator functions $g_N$ and $g_{N+1}$
is now obvious.

\begin{figure}
\centering
\includegraphics[width=2in]{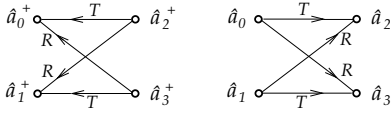}
\caption{The beam splitter described in the graphical method. In
the left graph we express the field operators $\hat{a}_0^\dagger$
and $\hat{a}_1^\dagger$ in respect with $\hat{a}_2^\dagger$ and
$\hat{a}_3^\dagger$ while in the right one we express $\hat{a}_2$
and $\hat{a}_3$ in respect with $\hat{a}_0$ and $\hat{a}_1$.}
\label{fig:graph_method_BS_inverse}
\end{figure}

\section{Applying the graphical method to a beam splitter and to a Mach-Zehnder interferometer}
\label{sec:graph_method_BS_MZI} In the following a novel,
graphical method allowing the fast computation of field operator
transformations will be introduced. In this section we apply this
method to a beam splitter and to a Mach-Zehnder interferometer.

\begin{figure}
\centering
\includegraphics[width=1.8in]{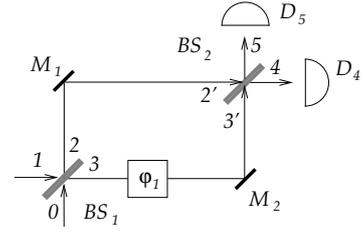}
\caption{The Mach-Zehnder interferometer. The delay ($\varphi_1$)
models the path length difference between the two arms of the
interferometer.} \label{fig:single_Mach_Zehnder_experiment}
\end{figure}

For a symmetrical, lossless beam splitter, the output annihilation
operators $\hat{a}_3$ and $\hat{a}_2$ can be written in respect
with the input field operators $\hat{a}_1$ and $\hat{a}_0$ and the
transmission ($T$) and reflection ($R$) coefficients (see
Fig.~\ref{fig:single_Mach_Zehnder_experiment}, beam splitter
$\text{BS}_1$) a result found in many textbooks (\emph{e.g.}
\cite{Lou03} page 214, \cite{Ger04} page 138). After some basic
manipulations, one easily obtains the input creation field
operators in respect with the output ones.

For the graphical method we start by representing the BS with the
butterfly-like graph, depicted in
Fig.~\ref{fig:graph_method_BS_inverse} (left graphic). The nodes
represent the fields in our points of interest and the arrows have
the amplitude coefficients that connect them. The field operator
$\hat{a}_0^\dagger$ is composed of two inverse paths from the
output, one from $\hat{a}_2^\dagger$ with an amplitude $T$ and one
from $\hat{a}_3^\dagger$ with an amplitude $R$ yielding
\begin{equation}
\hat{a}_0^\dagger=T\hat{a}_2^\dagger+R\hat{a}_3^\dagger
\end{equation}
and applying the same ideas to $\hat{a}_1^\dagger$, one quickly
obtains
\begin{equation}
\hat{a}_0^\dagger=R\hat{a}_2^\dagger+T\hat{a}_3^\dagger
\end{equation}
Similarly, from the right graph of
Fig.~\ref{fig:graph_method_BS_inverse}, the output field operators
$\hat{a}_2$ $\hat{a}_3$ can be obtained via two paths from the
input ports and, without any surprise, we obtain classical results
found in most textbooks.

Extension to the non-monochromatic case\footnote{Since we consider
narrowband non-monochromatic light, we assume our beam splitter to
be frequency-independent, \emph{i.e.} the coefficients $T$ and $R$
do not have a frequency dependence.} can be done using the
continuous frequency modes from Eq.~\eqref{eq:EX_fct_zeta_X_FREQ},
yielding
\begin{equation}
\tilde{\zeta_2}\left(\omega\right)\hat{a}_2
=T\tilde{\zeta_0}\left(\omega\right)\hat{a}_0+R\tilde{\zeta_1}\left(\omega\right)\hat{a}_1
\end{equation}
and
\begin{equation}
\tilde{\zeta_3}\left(\omega\right)\hat{a}_3
=R\tilde{\zeta_0}\left(\omega\right)\hat{a}_0+T\tilde{\zeta_1}\left(\omega\right)\hat{a}_1
\end{equation}
Denoting
$\hat{E}_k^{(+)}=\tilde{\zeta_k}\left(\omega\right)\hat{a}_k$ for
$k=2,3$ and performing the inverse Fourier transform of the above
equations one obtains the time-domain input-output relations.

For this simple example, the advantage of the graph-based method
is not at all obvious. It will, nonetheless, become important when
the devices discussed will become more complex.

A Mach-Zehnder interferometer (depicted in
Fig.~\ref{fig:single_Mach_Zehnder_experiment}) is composed of two
beam splitters and two mirrors. We denote again the input
(creation) field operators by $\hat{a}_0^\dagger$ and
$\hat{a}_1^\dagger$. The delay $\varphi_1$ introduced in the lower
path (\emph{i.e.} $\text{BS}_1$ -- $\text{M}_2$ -- $\text{BS}_2$)
accounts for the path length difference $\Delta{z}$ between the
two arms, \emph{i.e.} $\varphi_1=k\Delta{z}$, where $k$ is the
wavenumber. Being at a fixed frequency $\omega=k/c$, we have
$\varphi_1=\omega\tau_1$ where $\tau_1=\Delta{z}/c$.

\begin{figure}
\centering
\includegraphics[width=2.2in]{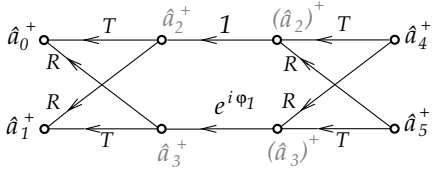}
\caption{The graph for the computation of the input operators
$\hat{a}_1^\dagger$ and $\hat{a}_0^\dagger$ in respect with the
output field operators $\hat{a}_4^\dagger$ and
$\hat{a}_5^\dagger$.
} \label{fig:graphical_method_direct_mzi_single_inv}
\end{figure}

In the case of the MZI, we construct the graph from
Fig.~\ref{fig:graphical_method_direct_mzi_single_inv}, composed of
two ``butterflies'' (for each BS) and the phase shift caused by
the path length difference of the two arms by the two horizontal
arrows connecting them with amplitudes\footnote{Strictly speaking,
we should model two delays, $\text{e}^{ikz_1}$ and
$\text{e}^{ikz_2}$, corresponding to the two paths. But since in
Quantum Mechanics absolute phases are irrelevant, we can set one
coefficient to $1$ and the other one to the phase contribution of
the path length difference.} $1$ and, respectively,
$\text{e}^{i\varphi_1}$. We model this delay through a positive
exponential factor (\emph{i.e.} $i\varphi_1=+i\omega\tau_1$) since
we are moving ``backwards in time'', from the input of
$\text{BS}_2$ to the output of $\text{BS}_1$. The names of the
intermediate field operators depicted in
Fig.~\ref{fig:graphical_method_direct_mzi_single_inv} are needed
only for explanatory purposes in the example below.

In order to illustrate the graphical method, we compute in detail
the field operator $\hat{a}_0^\dagger$ in function of the output
field operators $\hat{a}_4^\dagger$ and $\hat{a}_5^\dagger$. We
note that there are two possible paths connecting
$\hat{a}_4^\dagger$ to $\hat{a}_0^\dagger$: from
$\hat{a}_4^\dagger$ via $(\hat{a}'_2)^\dagger$ and
$\hat{a}_2^\dagger$ to $\hat{a}_0^\dagger$ with an amplitude $T^2$
and no phase shift and from $\hat{a}_4^\dagger$ via
$(\hat{a}'_3)^\dagger$ and $\hat{a}_3^\dagger$ to
$\hat{a}_0^\dagger$ with an amplitude $R^2$ and a phase shift
$\text{e}^{i\varphi_1}$. Another two paths connect
$\hat{a}_5^\dagger$ to $\hat{a}_0^\dagger$: from
$\hat{a}_5^\dagger$ via $(\hat{a}'_2)^\dagger$ and
$\hat{a}_2^\dagger$ to $\hat{a}_0^\dagger$ with an amplitude $TR$
and no phase shift and from $\hat{a}_5^\dagger$ via
$(\hat{a}'_3)^\dagger$ and $\hat{a}_3^\dagger$ to
$\hat{a}_0^\dagger$ with an amplitude $TR$ and phase shift
$\text{e}^{i\varphi_1}$. Summing up all these amplitudes takes us
to
\begin{equation}
\label{eq:single_MZI_a0_funct_of_a4_a5_graph}
\hat{a}_0^\dagger=\left(T^2+R^2\text{e}^{i\varphi_1}\right)\hat{a}_4^\dagger
+TR\left(1+\text{e}^{i\varphi_1}\right)\hat{a}_5^\dagger
\end{equation}
We obtained straight away this result by the simple inspection of
the graph from Fig.~\ref{fig:graph_method_BS_inverse}. Similarly,
writing down the amplitudes and phase shifts of the paths
connecting $\hat{a}_1^\dagger$ to $\hat{a}_4^\dagger$ and
$\hat{a}_5^\dagger$ yields
\begin{equation}
\label{eq:single_MZI_a1_funct_of_a4_a5_graph}
\hat{a}_1^\dagger=TR\left(1+\text{e}^{i\varphi_1}\right)\hat{a}_4^\dagger
+\left(T^2\text{e}^{i\varphi_1}+R^2\right)\hat{a}_5^\dagger
\end{equation}

\begin{figure}
\centering
\includegraphics[width=2.2in]{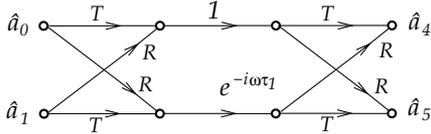}
\caption{The (frequency-domain) graph used for the computation of
the field operators $\hat{a}_4$ and $\hat{a}_5$ in respect with
the input operators  $\hat{a}_0$ and $\hat{a}_1$  for a
Mach-Zehnder interferometer.}
\label{fig:graphical_method_direct_single_mzi_a_FREQ}
\end{figure}

If we are interested in the operator functions connecting the
output annihilation field operators $\hat{a}_4$ and $\hat{a}_5$ to
the input field operators $\hat{a}_0$ and $\hat{a}_1$, we
construct the ``direct'' graph for monochromatic light, depicted
in Fig.~\ref{fig:graphical_method_direct_single_mzi_a_FREQ}. It is
composed of two ``butterflies'' (similar to the one depicted in
Fig.~\ref{fig:graph_method_BS_inverse}, right graphic) and a delay
line equivalent to a phase shift $\text{e}^{-i\omega\tau_1}$ since
we are at a fixed frequency. This time the exponent is
$-i\omega\tau_1$ since we are moving ``forward in time''. From the
graph depicted in
Fig.~\ref{fig:graphical_method_direct_single_mzi_a_FREQ} one can
readily write the four amplitudes connecting $\hat{a}_4$ to the
input field operators, having
\begin{equation}
\label{eq:single_MZI_a4_funct_of_a0_a1_graph}
\hat{a}_4=\left(T^2+R^2\text{e}^{-i\omega\tau_1}\right)\hat{a}_0
+TR\left(1+\text{e}^{-i\omega\tau_1}\right)\hat{a}_1
\end{equation}
and similarly for $\hat{a}_5$ yielding
\begin{equation}
\label{eq:single_MZI_a5_funct_of_a0_a1_graph}
\hat{a}_5=TR\left(1+\text{e}^{-i\omega\tau_1}\right)\hat{a}_0
+\left(T^2\text{e}^{-i\omega\tau_1}+R^2\right)\hat{a}_1
\end{equation}

\begin{figure}
\centering
\includegraphics[width=2.5in]{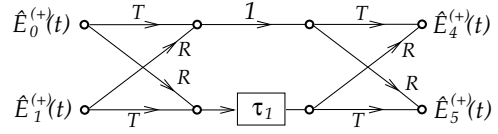}
\caption{The (time-domain) graph used for the computation of the
output electric field operators $\hat{E}^{(+)}_4\left(t\right)$
and $\hat{E}^{(+)}_5\left(t\right)$ in respect with the input
operators $\hat{E}^{(+)}_0\left(t\right)$ and
$\hat{E}^{(+)}_1\left(t\right)$.}
\label{fig:graphical_method_direct_single_mzi_E}
\end{figure}

We extend our analysis to continuous multi-mode non-monochromatic
light quanta by applying mode functions
$\tilde{\zeta}_0\left(\omega\right)$ and
$\tilde{\zeta}_1\left(\omega\right)$ to each input operator. We
end up with the input electric field operators
$\hat{E}^{(+)}_0\left(\omega\right)$ and
$\hat{E}^{(+)}_1\left(\omega\right)$. Nonetheless, we are
interested in a time-domain description, therefore we will perform
an inverse Fourier transform
$\hat{E}^{(+)}_k\left(\omega\right)\rightarrow\hat{E}^{(+)}_k\left(t\right)$.
Using Eq.~\eqref{eq:Fourier_DELAY_theorem} any phase shift will
become a time delay. We are now able to construct the graph from
Fig.~\ref{fig:graphical_method_direct_single_mzi_E}. The two beam
splitters are the same ``butterflies'' (assumed
frequency-independent), however the phase shift from
Fig.~\ref{fig:graphical_method_direct_single_mzi_a_FREQ} became a
time delay $\tau_1$, depicted in
Fig.~\ref{fig:graphical_method_direct_single_mzi_E} as a
rectangle.

One can express now the output field operators in respect with the
input ones. The operator $\hat{E}^{(+)}_4\left(t\right)$ can be
reached from $\hat{E}^{(+)}_0\left(t\right)$ via two paths
($T-1-T$ and $R-\text{delay }\tau_1-R$). Adding the two possible
paths from $\hat{E}^{(+)}_1\left(t\right)$ allows one to directly
write the result
\begin{eqnarray}
\hat{E}^{(+)}_4\left(t\right)=T^2\hat{E}^{(+)}_0\left(t\right)+R^2\hat{E}^{(+)}_0\left(t-\tau_1\right)
 \quad
\nonumber\\ 
 \quad
+TR\hat{E}^{(+)}_1\left(t\right)+TR\hat{E}^{(+)}_1\left(t-\tau_1\right)
\end{eqnarray}
and similarly, for the other output field operator
\begin{eqnarray}
\hat{E}^{(+)}_5\left(t\right)=TR\hat{E}^{(+)}_0\left(t\right)+TR\hat{E}^{(+)}_0\left(t-\tau_1\right)
 \quad
\nonumber\\ 
 \quad
+R^2\hat{E}^{(+)}_1\left(t\right)+T^2\hat{E}^{(+)}_1\left(t-\tau_1\right)
\end{eqnarray}

\section{Applying the graphical method to a double Mach-Zehnder interferometer}
\label{sec:graph_method_2MZI} 
\begin{figure}
\centering
\includegraphics[width=2.5in]{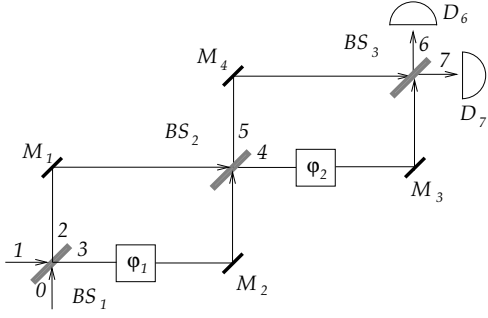}
\caption{The experiment proposed in Section
\ref{sec:graph_method_2MZI}. The first Mach-Zehnder interferometer
is composed of the beam splitters $\text{BS}_1$ and $\text{BS}_2$
together with the mirrors $\text{M}_1$ and $\text{M}_2$.
Similarly, the second Mach-Zehnder interferometer is composed of
$\text{BS}_2$ , $\text{BS}_3$, $\text{M}_3$ and $\text{M}_4$. The
delays ($\varphi_1$ and $\varphi_2$) model the path length
difference in each MZI.}
\label{fig:double_Mach_Zehnder_experiment}
\end{figure}

We introduce and discuss in the
following a double MZI setup, depicted in
Fig.~\ref{fig:double_Mach_Zehnder_experiment}. The beam splitters
$\text{BS}_1$ and $\text{BS}_2$ with the two mirrors $\text{M}_1$
and $\text{M}_2$ form the first MZI. The delay $\varphi_1$ models
the path length difference. The second MZI is composed of the beam
splitters $\text{BS}_2$ and $\text{BS}_3$, together with the two
mirrors $\text{M}_3$ and $\text{M}_4$. Similarly, the delay
$\varphi_2$ models the path length difference. It is assumed that,
with the corresponding delays taken out of the experiment, each
MZI has equal length arms. The photo-detectors $D_6$ and $D_7$
(assumed ideal) are installed at the two outputs of $\text{BS}_3$.

Results concerning the double Mach-Zehnder interferometer have
been discussed in detail in \cite{Ata14} and will be used in the
following as reference.

\begin{figure}
\centering
\includegraphics[width=2.8in]{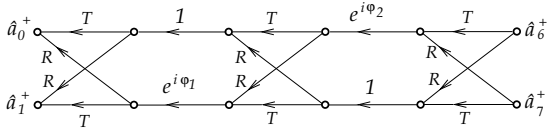}
\caption{The (frequency domain) graph used to compute the input
field operators $\hat{a}_0^\dagger$ and $\hat{a}_1^\dagger$ in
respect with the output field operators $\hat{a}_6^\dagger$ and
$\hat{a}_7^\dagger$ for a double Mach-Zehnder interferometer.}
\label{fig:graphical_method_direct_mzi_double_inv}
\end{figure}

Similar to the previous case studies, we construct a graph,
depicted in Fig.~\ref{fig:graphical_method_direct_mzi_double_inv}.
Each beam splitter is depicted as a butterfly (with coefficients
$T$ and $R$) and the path length difference in each MZI is
modelled as a phase shift. We start by expressing the field
operator $\hat{a}_0^\dagger$ in respect with the output field
operators. From
Fig.~\ref{fig:graphical_method_direct_mzi_double_inv} one can see
that there are four possible paths from $\hat{a}_6^\dagger$ to
$\hat{a}_0^\dagger$ yielding the amplitudes:
$T^3\text{e}^{i\varphi_2}$, $TR^2$,
$TR^2\text{e}^{i(\varphi_1+\varphi_2)}$ and
$TR^2\text{e}^{i\varphi_1}$.  Likewise, there are four paths from
$\hat{a}_7^\dagger$ to $\hat{a}_0^\dagger$. Therefore, the input
operator $\hat{a}_0^\dagger$ in respect with the output operators
$\hat{a}_6^\dagger$ and $\hat{a}_7^\dagger$ can be directly
written yielding
\begin{eqnarray}
\label{eq:double_MZI_a0_dagger_funct_of_a6_a7_graph}
\hat{a}_0^\dagger
=\left(T^3\text{e}^{i\varphi_2}+TR^2+TR^2\text{e}^{i(\varphi_1+\varphi_2)}+TR^2\text{e}^{i\varphi_1}\right)\hat{a}_6^\dagger
\nonumber\\
+\left(R^3\text{e}^{i(\varphi_1+\varphi_2)}+T^2R\text{e}^{i\varphi_1}+T^2R\text{e}^{i\varphi_2}+T^2R\right)\hat{a}_7^\dagger\quad
\end{eqnarray}
Similar arguments allow us to directly  write the input operator
$\hat{a}_1^\dagger$ in respect with the output ones,
\begin{eqnarray}
\label{eq:double_MZI_a1_dagger_funct_of_a6_a7_graph}
\hat{a}_1^\dagger
=\left(T^2R\text{e}^{i\varphi_2}+R^3+T^2R\text{e}^{i(\varphi_1+\varphi_2)}+T^2R\text{e}^{i\varphi_1}\right)\hat{a}_6^\dagger
\nonumber\\
+\left(T^3\text{e}^{i\varphi_1}+TR^2\text{e}^{i(\varphi_1+\varphi_2)}+TR^2\text{e}^{i\varphi_2}+TR^2\right)\hat{a}_7^\dagger\quad
\end{eqnarray}
Eqs.~\eqref{eq:double_MZI_a0_dagger_funct_of_a6_a7_graph} and
\eqref{eq:double_MZI_a1_dagger_funct_of_a6_a7_graph} could have
been found using the traditional, iterative method \cite{Ata14}.
The graphical method, however, gave these results faster and in a
more intuitive way.

\begin{figure}
\centering
\includegraphics[width=2.9in]{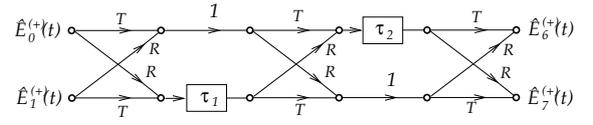}
\caption{The (time-domain) graph for the computation of the output
operators $\hat{E}^{(+)}_6\left(t\right)$ and
$\hat{E}^{(+)}_7\left(t\right)$ in respect with the input ones,
$\hat{E}^{(+)}_0\left(t\right)$ and
$\hat{E}^{(+)}_1\left(t\right)$.}
\label{fig:graphical_method_direct_mzi_double_fwd_TIME}
\end{figure}

Extension to non-monochromatic light quanta can be done using the
same path as before. Therefore, we construct the graph from
Fig.~\ref{fig:graphical_method_direct_mzi_double_fwd_TIME}. One
can express now the output field operator
$\hat{E}^{(+)}_6\left(t\right)$ in respect with the input electric
field operators $\hat{E}^{(+)}_0\left(t\right)$ and
$\hat{E}^{(+)}_1\left(t\right)$ by simply inspecting the graph. We
have four possible paths connecting
$\hat{E}^{(+)}_0\left(t\right)$ to $\hat{E}^{(+)}_6\left(t\right)$
yielding the amplitudes and delays: $T^3$ with a delay of
$\tau_2$, $TR^2$ with a delay of $\tau_1$, $TR^2$ with no delay
and finally $R^3$ with a delay of $\tau_1+\tau_2$. Adding the
contribution from $\hat{E}^{(+)}_1\left(t\right)$ yields the final
expression
\begin{eqnarray}
\label{eq:E6_plus_fct_E0_E1_General_TR_GRAPH}
\hat{E}^{(+)}_6\left(t\right)
=T^3\hat{E}^{(+)}_0\left(t-\tau_2\right)+TR^2\hat{E}^{(+)}_0\left(t-\tau_1\right)\quad
\nonumber\\ 
+TR^2\hat{E}^{(+)}_0\left(t-\tau_1-\tau_2\right)
+TR^2\hat{E}^{(+)}_0\left(t\right)
\nonumber\\ 
+T^2R\hat{E}^{(+)}_1\left(t-\tau_1\right)
+T^2R\hat{E}^{(+)}_1\left(t-\tau_1-\tau_2\right)
\nonumber\\ 
+T^2R\hat{E}^{(+)}_1\left(t-\tau_2\right)+R^3\hat{E}^{(+)}_1\left(t\right)
\end{eqnarray}
Considering the four paths from $\hat{E}^{(+)}_0\left(t\right)$
and the other four from $\hat{E}^{(+)}_1\left(t\right)$ to
$\hat{E}^{(+)}_7\left(t\right)$ one finds
\begin{eqnarray}
\label{eq:E7_plus_fct_E0_E1_General_TR_GRAPH}
\hat{E}^{(+)}_7\left(t\right)
=T^2R\hat{E}^{(+)}_0\left(t-\tau_2\right)+T^2R\hat{E}^{(+)}_0\left(t-\tau_1\right)
\nonumber\\ 
+T^2R\hat{E}^{(+)}_0\left(t\right)+R^2\hat{E}^{(+)}_0\left(t-\tau_1-\tau_2\right)
\nonumber\\ 
+T^3\hat{E}^{(+)}_1\left(t-\tau_1\right)
+TR^2\hat{E}^{(+)}_1\left(t-\tau_1-\tau_1\right)
\nonumber\\ 
+TR^2\hat{E}^{(+)}_1\left(t\right)+TR^2\hat{E}^{(+)}_1\left(t-\tau_2\right)
\end{eqnarray}
Eqs.~\eqref{eq:E6_plus_fct_E0_E1_General_TR_GRAPH} and
\eqref{eq:E7_plus_fct_E0_E1_General_TR_GRAPH} were also obtained
with merely a visual inspection of the graph depicted in
Fig.~\ref{fig:graphical_method_direct_mzi_double_fwd_TIME}. They
are identical to the results from \cite{Ata14} but the effort in
obtaining them was much lower.

\section{Conclusions}
\label{sec:conclusions} In this paper we introduced and discussed
a graphical method allowing the easy computation of field operator
transformation for linear lossless devices in quantum optics
comprising beam splitters and interferometers. Besides the
advantage in the speed of calculation, this method offers an
intuitive physical interpretation: operators transform via a sum
of probability amplitudes from each available path in the
considered optical system. Direct and inverse graphs can be built
in function of the required operator equation. Extension to the
non-monochromatic case is done via time-domain graphs, where the
output electric field operators can be computed in respect with
the input ones in the same, intuitive, graphical manner.

%
%

\end{document}